\documentclass{aa}
\usepackage[dvips,xdvi]{graphics}

\begin{document}
\newcommand{\MIC}{\mbox{$\mu$m}}
\newcommand{\MOL}{\mbox{$M/L{\rm _H}$}}
\newcommand{\HA}{\mbox{H$\alpha$}}
\newcommand{\EW}{\mbox{$W_\lambda$}}
\newcommand{\LSUN}{\mbox{L$_\odot$}}
\newcommand{\COspec}{\mbox{$\rm[CO]_{spec}$}}
\newcommand{\COphot}{\mbox{$\rm[CO]_{phot}$}}
\newcommand{\CO}{\mbox{[CO]}}
\newcommand{\CANOM}{\mbox{[C/Fe]}}
\newcommand{\Teff}{\mbox{$T_{\rm eff}$}}
\newcommand{\Mbol}{\mbox{$M_{\rm bol}$}}
\newcommand{\Lbol}{\mbox{$L_{\rm bol}$}}

\title{Is the [CO] index an age indicator for star forming galaxies?}

\author
{
L. Origlia\inst{1}
\and
E. Oliva\inst{2}
}
          
\institute{ 
Osservatorio Astronomico di Bologna, Via Ranzani 1, I--40127 Bologna, Italy
\and
Osservatorio di Arcetri, Largo E. Fermi 5, I-50125 Firenze, Italy
}

\offprints{L. Origlia, e-mail origlia@astbo3.bo.astro.it }

\thesaurus{ 03 (
               11.19.3;          
               11.19.4;          
               11.19.5;          
               13.09.6          
               )
          }

\date{ Received Oct 29 1999; Accepted Mar 6 2000}
\maketitle

\begin{abstract} 
The classical [CO] index, 
i.e. the strength of the $\Delta v$=2 CO absorption bands
starting at 2.29 \MIC, is sometimes used to constrain the maximum age
of star formation events in galaxies.
In this paper we critically analyze
\begin{description}
\item[{\it i)\hglue3pt}] 
theoretical models which could predict either a factor of $>$2 drop
or a pronounced increase
of [CO] at ages older than 100 Myr, depending on the evolutionary tracks
one adopts (see Fig.~\ref{evolco}).
\item [{\it ii)}] 
observational data for young clusters in the 
LMC
which do not show any strong relationship between the CO index and cluster
age
(see Fig.~\ref{figobs}).
\end{description}
The above scenario indicates that the value of [CO] does not 
provide a reliable tool
for estimating the age of stellar populations older than $\sim$10 Myr,
i.e. after the first red supergiants have been formed.
%
The contradictory results of theoretical models reflect problems
in treating the evolution along the Asymptotic Giant Branch (AGB).
In particular, those evolutionary synthesis models using stellar tracks 
which do not include the thermal pulsing AGB phase produce 
too weak CO features at 100--1000 Myr, i.e. in the range of ages when
the near infrared emission is
dominated by thermal pulsing AGB stars.
\end{abstract}


\section{Introduction} 

An ideal instantaneous burst of star formation generates a so--called 
Simple Stellar Population (SSP),
that is  a stellar system which is coeval and initially 
chemically homogeneous (see Renzini \& Buzzoni \cite{buzzoni86}). 
The integrated near IR luminosity of a SSP is dominated by 
red stars since its very early stage of evolution ($\simeq$10 Myr), when 
massive stars ($<$40~M$_{\odot}$) evolve as red supergiants. 
When the stellar system gets older ($\ga$100 Myr) intermediate 
mass giants evolving along the AGB and, 
after a few Gyr, low mass giants near the tip of the Red Giant Branch 
(RGB) dominate the integrated IR and bolometric luminosities
(e.g. Renzini \& Buzzoni \cite{buzzoni86}, 
Chiosi et al. \cite{chiosi86}).
The time evolution of the observable parameters related to a
 SSP, such as e.g. photometric colours and spectral indices, 
provides the basic ingredient for constructing 
evolutionary models of star forming galaxies.

Among the photometric and spectroscopic indices used to study the 
red stars of a SSP,
the CO index has attracted quite some attention
as a potential tool to trace red supergiants,
i.e. young stellar systems.
This idea primarily derives from the fact that field stars of similar spectral
types show different CO indices depending
on their spectral class,
the strongest features being found in supergiants
(see e.g. Fig. 4 of  Kleinmann \& Hall 1986, hereafter \cite{KH86}).
Several attempts of predicting the evolution of the CO index of a SSP
appeared in the literature and were applied to the interpretation
of IR spectral observations of starburst galaxies 
(e.g. Doyon et al. \cite{doyon94},
Shier et al. \cite{shier96}, Goldader et al. \cite{goldader97},
Mayya \cite{mayya97}, Leitherer et al. \cite{leitherer99}).
Most of the models are restricted to solar metallicities
and predict a pronounced maximum at $\simeq$10 Myr
followed
by a quite rapid and steady decline. The CO index drops by almost a factor
of 3 at $\simeq$100 Myr and, noticeably, reaches values much lower than those
observed in old ($\ga$10 Gyr) Galactic
globular clusters and spheroidal galaxies
of quasi--solar metallicities. 
The few models at sub--solar metallicities predict a similar time
evolution with shallower CO features at all epochs.

Taken at face value, these models would imply that star forming galaxies
with prominent CO absorption features must be dominated by a
young ($<$100 Myr) star formation event, while more mature,
but still relatively young systems of a few $\times\;$100 Myr should 
be characterized by quite
weak CO absorption features. In other words, the CO index could provide
a powerful tool to constrain the age of the major star formation event
of galaxies.

This paper is a critical re--analysis of the time evolution of the CO
index in simple SSPs and star forming galaxies. In Sect.~\ref{CO1} we
describe
the, sometimes confusing, definition of CO index and discuss its relationship
with stellar parameters.
In Sect.~\ref{model_evol} we present theoretical curves based 
on different stellar 
evolutionary models and briefly discuss the possible reasons for their
very different behaviours.
In Sect.~\ref{data} we compare the predicted evolution of [CO] with
measured parameters of template stellar clusters in the Magellanic Clouds,
old globular clusters in the Galaxy, normal and starburst galaxies.
In Sect.~\ref{conclusions} we draw our conclusions.

\section{The CO index} 
\label{CO1}

\subsection{Spectroscopic and photometric definitions. }
The CO index was originally defined as the magnitude difference
between a relatively narrow filter ($\Delta\lambda\!\simeq\!0.1$ \MIC)
centered at 2.3 \MIC, which includes 
the first four band--heads of $\Delta v$=2 CO roto--vibrational transitions,
and a similarly narrow filter centered at 2.2 \MIC\  (Baldwin et al.
\cite{baldwin73}). The central wavelength of 
the CO filter was then increased to 2.36 \MIC\  and slightly different
filter parameters were adopted by different groups.
A comprehensive database of CO photometric
measurements was produced in the 70--80's. These include measurements of
field stars (e.g. McWilliam \& Lambert \cite{lambert84}), Galactic
globular clusters (Frogel et al. \cite{frogel83}),
young stellar clusters in Magellanic Clouds (Persson et al. \cite{persson83})
and old spheroidal galaxies (Frogel et al. \cite{frogel78}).
These data are still considered a fundamental benchmark for verifying
the predictions of stellar evolutionary models.

The spectroscopic CO index was defined by
\cite{KH86} who measured 
the (2,0) band--head at 2.29 $\mu$m from medium resolution 
($\lambda/\Delta\lambda\!\simeq\!2500$) spectra of a sample
of field stars. The strength of this band is unequivocally 
defined as the ratio between the fluxes integrated over 
narrow wavelength ranges centered on the line and nearby
continuum, i.e. 2.2924--2.2977  and 2.2867--2.2919 \MIC,
and expressed in terms of magnitudes.
The {\it same quantity} is sometimes given in terms of equivalent width
(e.g. Origlia et al.  \cite{origlia93})
and the numbers are simply related by 
$$ \COspec = -2.5\ \log 
\left(1 - {\EW(2.29)\over 53\, {\rm \AA} }\right) \ \ \ \  {\rm mag} \eqno(1) $$
where \COspec\  is the spectroscopic index and \EW(2.29) is the equivalent
width of the (2,0) band--head. 

The relationship between spectroscopic and photometric indices is not
obvious because they are based on measurements of different quantities
which have different behaviours on the stellar physical parameters,
e.g. \COphot\   also depends on the $^{12}$C/$^{13}$C ratio
(see McWilliam \& Lambert \cite{lambert84}).
Nevertheless, an empirical correlation between the two indices is generally
adopted following from observations of giant stars in the field. This yields
(see Fig.~3 of \cite{KH86})
$$ \COphot \simeq 0.57 \,\COspec - 0.01   \ \ \ \ \ \  {\rm mag} \eqno(2) $$
%

%
To complicate the scenario further, 
other definitions of the 
spectroscopic index exist in the literature. These are based on 
spectroscopic measurements
of equivalent widths over wavelength ranges much broader 
than those used by \cite{KH86} and similar to that adopted in
the photometric definition. 
The most popular of these intermediate indices is that proposed by 
Doyon et al.  (\cite{doyon94}) which measures the equivalent width
over the 2.31--2.40 $\mu$m range relative to a continuum which
is extrapolated from shorter wavelengths.
The specific advantage of this definition is that it allows measurements
of \CO\  even from spectra of relatively poor quality, while
proper measurement of the spectroscopic index requires high s/n spectra 
with resolution $R\!\ga\!1000$.

These ``broad spectral indices'' are calibrated using measurements of
field stars and can be therefore converted to photometric indices
using empirical relationships similar to Eq.~(2). 
To avoid confusion we will hereafter express measured and predicted
quantities in terms of \COphot\  
with additional comments on how
it was obtained or scaled from spectroscopic measurements.

\subsection{ CO index and stellar parameters}
\label{CO_and_stellar_param}

In general, the strength of the CO features depends on the following
stellar parameters (for a more detailed discussion
see Sect. 4.1 of Origlia et al. \cite{origlia93})
\begin{description}
\item[--\ ] Effective temperature, \Teff, 
 which sets the CO/C relative abundance.
\item[--\ ] Surface gravity, $g$, which influences the H$^-$/H equilibrium
and hence modifies the total gas column density of the
photosphere. In cool stars with CO/C$\simeq$1 the column density of CO
scales as $\approx g^{-1/3}$ and, therefore, the lines become more
opaque when the gravity decreases.
\item[--\ ] Microturbulent velocity, $\xi$, which determines the Gaussian
width of the lines and, therefore, the equivalent width of saturated
lines (the CO transitions are semi-\-forbidden
and have very weak Lorentzian wings).
\item[--\ ] Metallicity and carbon relative abundance, which define
the value of C/H. For a given set of \Teff\  and $g$, all CO line 
opacities scale linearly with the carbon abundance C/H.
\end{description}
The well known correlation between CO index and spectral type of giant stars
is due to a complex combination of the effects of the first three parameters.
The variation of \Teff\  
is important only up to early K stars where most of the carbon
is already in the form of CO. Therefore, in later spectral types the variation
of \Teff\  has virtually no direct effect on [CO], 
i.e. {\it the CO index is not a thermometer for cool stars}.

The steady deepening of the CO features 
along the K4$\,$III--M7$\,$III sequence 
is driven by the decrease of surface gravity and 
increase of microturbulent velocity. The variation of surface gravity,
$\Delta\log\,g\!\simeq\!-0.8$ from early K to late M giants
(McWilliam \& Lambert \cite{lambert84}),
follows from the fact that field red giants
are stars with similar mass and quasi--solar metallicities taken
at different phases of their evolution on the RGB.
Thus M$\,$III stars, being cooler and more luminous, have a 
lower surface gravity than K giants.

The value of $\xi$ cannot be directly related to the stellar parameters
but, based on detailed spectroscopic observations,
is found to increase when the bolometric luminosity increases.
To a first approximation, $\xi$ scales linearly with $\log(\Lbol)$
and, therefore, late M giants have microturbulent
velocities about 0.8 km/s higher than
early K$\,$III stars (see e.g. Tsuji \cite{tsuji86}, \cite{tsuji91}). 

The effect of microturbulent velocity becomes particularly prominent
in red supergiants and accounts for the fact that 
class I stars have much stronger [CO] than giants of similar temperatures and
gravities (see e.g. Tsuji et al. \cite{tsuji94}, McWilliam \& Lambert 
\cite{lambert84}).
The derived values of $\xi$ scale $\simeq$ linearly
with $\log(\Lbol)$, i.e. a behaviour similar to that found in giant stars.
%
Therefore, for practical purposes, the variation of \CO\  
in red giants and supergiants can be reproduced 
adopting an empirical, linear relationship between $\xi$ 
and bolometric luminosity, namely:
$$ \xi \simeq 2.0-0.4\,\Mbol \ \ \ \ \ \ \ {\rm km/s} \eqno(3) $$
An indirect test (and confirmation) of this equation comes from
integrated spectral analysis of old and metallic stellar systems
(Origlia et al.~\cite{origlia97}) and young clusters in Magellanic
Clouds (Oliva \& Origlia~\cite{oliva98}). The derived values
of average microturbulent velocities, namely $\overline\xi\!\simeq\!2$
and $\overline\xi\!\simeq\!4-5$ km/s for old and young systems, respectively,
are in good agreement with those derived from spectral synthesis
models adopting Eq.~(3).\\

The effect of carbon abundance is relatively unimportant in field stars,
which span a relatively narrow range of metallicities, but becomes evident
in stellar clusters of low metallicity which are characterized by very
weak CO features (see the left panel of Fig.~\ref{figobs}).
However, it should be kept in mind that
the nice correlation between [CO] and metallicity  of Fig.~\ref{figobs} 
also reflects the metallicity dependence of the temperature of giant stars,
i.e. lower metallicity clusters have weaker [CO] not only because their
giant stars have less carbon, but also because they are warmer than 
those in more metallic stellar systems
(e.g. Origlia et al. \cite{origlia97}).

\begin{figure}
\centerline{\resizebox{8.8cm}{!}{\rotatebox{0}{\includegraphics{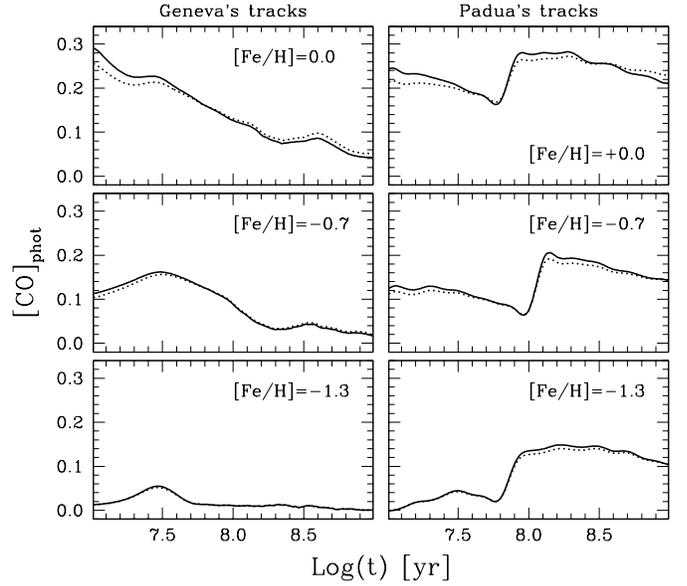}}}}
\caption{
Predicted evolution of the CO index for different metallicities and
using different stellar evolutionary models. The left panels
are based on the Geneva's tracks (Schaller et al. \cite{schaller92},
Charbonnel et al. \cite{charbonnel93}, Schaerer et al. \cite{schaerer93})
while those in the right hand panels are based on the Padua's tracks 
(Bertelli et al. \cite{bertelli94}). 
The solid lines represent our `reference models' which 
are computed adopting an empirical relationship between 
microturbulent velocity and bolometric luminosity, i.e. using Eq.~(3).
For comparison, the dotted curves show the results obtained by assuming
a constant value of microturbulent velocity $\xi$=4.0
(see Sects.~\ref{CO_and_stellar_param},\ref{model_evol} for details).
Note the very different behaviours beyond $\sim$100 Myr, i.e. when the near IR
emission is dominated by stars evolving on the AGB. This simply reflects
the different extent of the AGB in the two sets of models 
(see Fig.~\ref{hr}).
}
\label{evolco}
\end{figure}

\begin{figure}
\centerline{\resizebox{8.8cm}{!}{\rotatebox{0}{\includegraphics{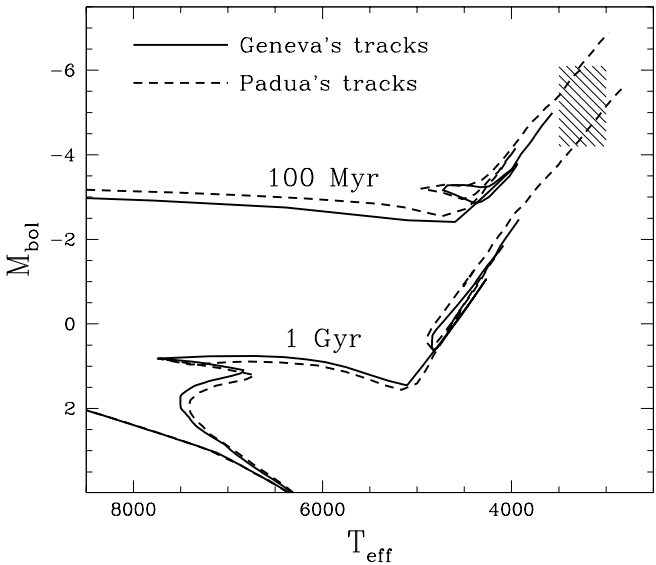}}}}
\caption{
HR diagram with theoretical isochrones from
different stellar evolutionary models, all curves are
for solar metallicities. 
The solid lines
are based on the Geneva's tracks (Schaller et al. \cite{schaller92},
Charbonnel et al. \cite{charbonnel93}, Schaerer et al. \cite{schaerer93})
which stop the computation at the onset of thermal pulses
and, therefore, cuts the AGB  at unrealistically
high temperatures and low luminosities.
The dotted curves are from Bertelli et al. (\cite{bertelli94}) which
follow the evolution through the thermal pulse phase and hence
produce more luminous and cooler AGB stars. However, the 
extent of the AGB is probably overestimated by these models
which assume quite low mass loss rates along the AGB 
(see Sect.~\ref{model_evol}).
\hfill\break
The shaded region show the loci of AGB stars in intermediate age Magellanic
Cloud clusters (Aaronson \& Mould \cite{aaronson82}, Frogel et al. 
\cite{frogel90}).
}
\label{hr}
\end{figure}

\section{Modelling the evolution of [CO] in a SSP}
\label{model_evol}

The integrated CO index of a SSP with a given metallicity and
age can be most conveniently determined from the integrated synthetic spectrum
using the same methods adopted for measuring the index from the observational
data. 
%
The integrated synthetic spectrum can be expressed as
$$ F_\lambda = \int \phi(M)\, L_{\rm K}(M)\,
     f_\lambda(T,g,\xi,\CANOM) \, dM  \eqno(4) $$
where 
$T$=$T(M)$, $g$=$g(M)$ and $L_{\rm K}(M)$ are
the stellar temperature, gravity and monochromatic luminosity 
\footnote{ $L_{\rm K}$ (erg s$^{-1}$ $\mu$m$^{-1}$) 
is the luminosity per $\lambda$-unit at 2.2 $\mu$m}
along the isochrone. These parameters are explicitly tabulated
by the Padua's models (Bertelli et al. \cite{bertelli94})
while can be derived from the Geneva's evolutionary tracks 
as outlined in Leitherer et al. (\cite{leitherer99}).

The shape of the initial mass function, $\phi(M)$, has only a minor effect
on the derived indices and, for all practical purposes, can
be approximated by a standard Salpeter's IMF 
($\phi(M)\!\propto\!M^{-2.35}$).
The function $f_\lambda$ is the spectrum (normalized to unity at 2.2
$\mu$m) appropriate for
a star of the given metallicity, temperature and gravity.
This function is often evaluated using observations of field stars whose
spectral types are converted into surface temperatures and luminosities
using empirical calibrations.
However, this method is limited to stars with
quasi--solar metallicity and is particularly uncertain for supergiants
and AGB stars.

A more objective estimate of $f_\lambda$ can be obtained from
model stellar atmospheres which yield a synthetic spectrum
with the desired resolution 
for a given set of metallicity ($Z$), effective temperature ($T$), 
surface gravity ($g$), 
microturbulent velocity ($\xi$) and carbon relative abundance (\CANOM). 
The last two quantities should be treated as free parameters because they
cannot be unequivocally related to other physical
parameters of the star. 
A detailed description of the procedure used to construct the 
synthetic spectra can be found in Origlia et al. (\cite{origlia93}).
 
The time evolution of the CO index simply follows from Eq.~(3) using
theoretical isochrones at different ages.
The results are plotted in Fig.~\ref{evolco} where we also evaluate the
effect of using different stellar evolutionary models. 
These agree in predicting strong \CO\  
at early times, when the near IR emission is dominated
by red supergiants (i.e. $t\!\la\!100$~Myr). Indeed, the comparison
with observations is not yet satisfactory because models predict
too warm red supergiants, especially at sub--solar metallicities
(e.g. Oliva \& Origlia \cite{oliva98}, Origlia et al.~\cite{origlia99}).

The most striking result in Fig.~\ref{evolco} is the very different 
behaviour in the AGB phase (i.e. $t\!\ga\!100$~Myr) where the
predicted CO index varies by large ($>$3) factors depending on the
adopted stellar evolutionary tracks. In particular, the steady
decay of [CO] in the curves based on the Geneva's models contrasts
with the increase at the onset of the AGB predicted by the
Padua's tracks. Interestingly, the latter predict strong CO at
low metallicities where the other models give [CO]$\simeq$0.

To investigate the reason(s) for the very different behaviours in the
AGB phase, it is instructive to compare the model isochrones
which are displayed in Fig.~\ref{hr} for two representative
ages at solar metallicity. The Geneva's curves stop at relatively
low luminosities and high temperatures ($>$3600 K), this implies that the
stars dominating the IR emission are warm enough to dissociate CO,
have quite large surface gravities and relatively low luminosity. 
Thus the CO bands are
weak because the CO/C relative abundance, the column density of the
photosphere and the microturbulent velocity are all relatively small
(see Sect.~\ref{CO_and_stellar_param}).
The AGB in the Padua's models, on the contrary, extends to very high
luminosities and low temperatures. This implies low surface gravities
and large microturbulent velocities, i.e. deep CO features.

The different extent of the AGB in Fig.~\ref{hr} primarily follows
from assumptions made by the models. The Geneva's tracks arbitrarily
stop at the onset of thermal pulses, while the Padua's computations
follow this phase up to the very end of the double shell burning
using semi--analytical approximations.
However, both approaches are unrealistic because the  
evolution along the AGB, which for sure extends well into the 
thermal pulses phase, 
is regulated and shortened by the strong mass--loss experienced by AGB stars,
a parameter not included in the Bertelli et al. (\cite{bertelli94}) tracks.
Moreover, one should keep in mind that the predicted stellar temperatures 
are also quite uncertain and, 
probably, too warm (Chieffi et al. \cite{chieffi95}, Oliva \& Origlia 
\cite{oliva98}).
This effect is visible in Fig.~\ref{hr} where one can notice that,
within the part of the AGB covered by both models,
the Padua's tracks are systematically warmer than those of Geneva.
Indeed, recent developments of the theory of convective energy transfer
indicate that all the temperatures of red stars in the Padua's tracks
should be decreased by 200--300 K (Bressan, private communication).

Therefore, the ``true'' AGB is likely to be less extended but redder than
predicted by the Padua's tracks and the two effects on the [CO]
should compensate each others.
In practice, the ``true'' CO index during the AGB phase 
is probably similar to that plotted in the right hand
panels of Fig.~\ref{evolco}. 
For the moment being,
assuming [CO]$\,\simeq\,$constant with time is probably a fair 
approximation which, at least, does not lead to far reaching conclusions
on the age of stellar systems.

\begin{figure}
\centerline{\resizebox{8.8cm}{!}{\rotatebox{0}{\includegraphics{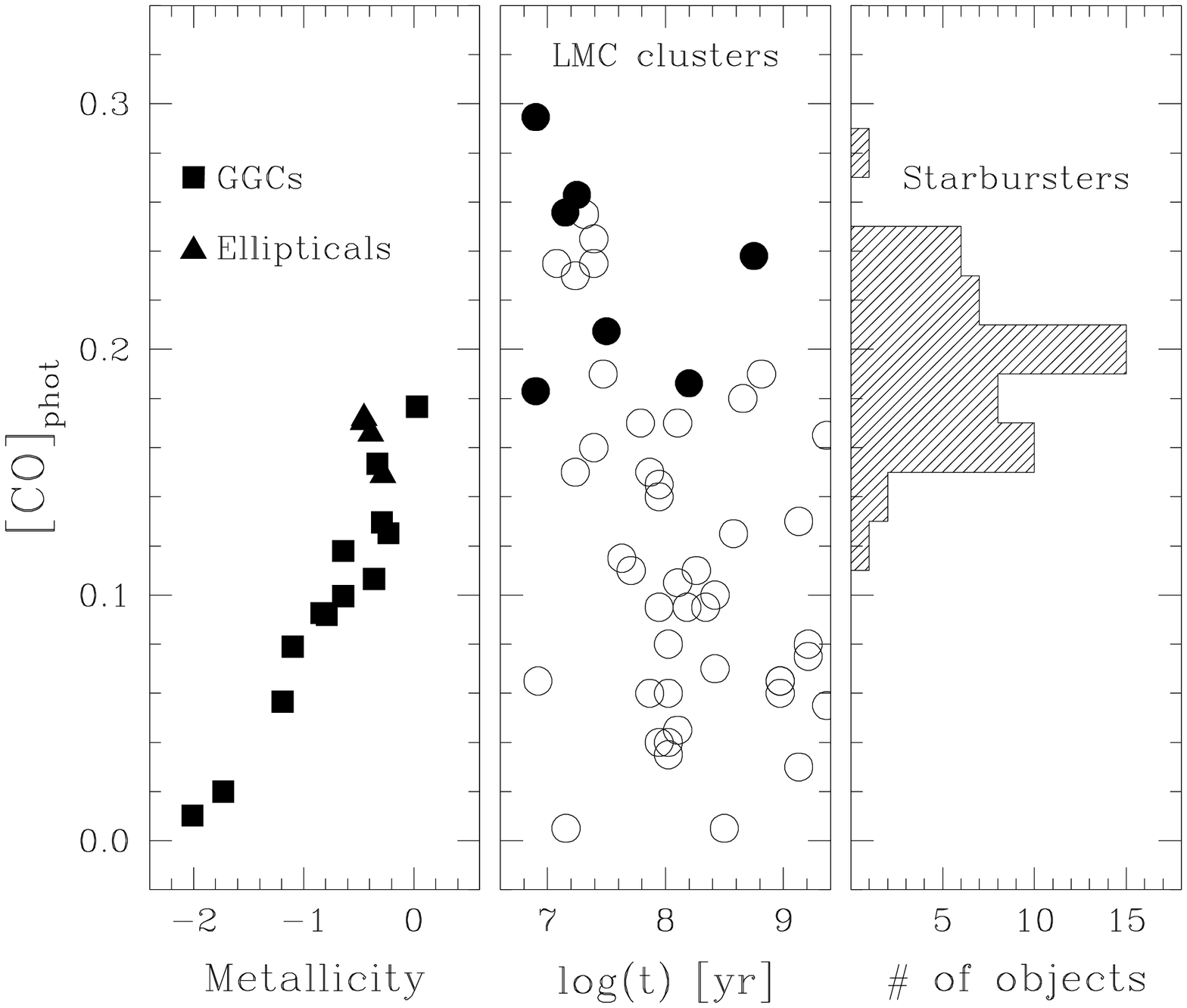}}}}
\caption{
Observed [CO] in different 
samples of stellar clusters and galaxies.
The first panel includes old Galactic globular clusters 
(Origlia et al. \cite{origlia97}, equivalent widths
scaled to \COphot\   using Eqs. (1) and (2)),
note the quite tight correlation between CO index and metallicity.
\hfil\break
The second plot includes younger clusters in the LMC,
open circles are from
the photometric data of Persson et al. (\cite{persson83}) while
the filled symbols are from the spectroscopic equivalent widths
of Oliva \& Origlia (\cite{oliva98}), scaled to \COphot\   using
Eqs. (1) and (2).
The large scatter of the points at a given age
reflects variations of metallicities, statistical effects
related to the intrinsic small number of luminous red stars in the clusters
and, possibly, field contamination
(see also Sect.~\ref{data}).
\hfil\break
The third panel shows the distribution of the CO indices observed in
starburst galaxies. The data are from Oliva et al. 
(\cite{oliva95}, \cite{oliva99},
equivalent widths transformed into \COphot\
using Eqs. (1),(2)) and from
Shier et al. (\cite{shier96}), Goldader et al. (\cite{goldader97})
and Doyon et al. (\cite{doyon94}) whose ``broad
spectroscopic index'' is translated into \COphot\   using the prescriptions
in the latter reference.
Note that virtually all of the starbursters are within the range
covered by $10^7$--$10^9$ yr old LMC clusters. 
}
\label{figobs}
\end{figure}

\section{Available data and SSP templates}
\label{data}

Stellar clusters in the Large Magellanic Cloud
provide a convenient and, in practice, unique set of templates for young
and intermediate age stellar populations spanning a wide range of 
metallicities (between 1/100 and $\simeq$ solar, e.g. 
Sagar \& Pandey \cite{sagar89}).
%
The available [CO] data are summarized in 
Fig.~\ref{figobs}. The large scatter of the points at a given age
reflects variations of metallicities, statistical effects
related to the intrinsic small number of luminous red stars in the clusters
and, possibly, field contamination (see e.g. Chiosi et al.~\cite{chiosi86},
Santos et al.~\cite{santos}).
A further complication occurs beyond $\approx$600~Myr, when carbon
stars appear. These often display a very red spectrum with strong continuum
emission from the envelope which dilutes the CO bands and produce an
anti--correlation between J--K colours and [CO] index (Persson et al. 
\cite{persson83}).
Note in particular that
the mild anti--correlation between CO index and age visible in 
Fig.~\ref{figobs}
also reflects the fact that the older clusters are, on average,
less metallic than the youngest one's. 

For the purposes of this paper, the most important fact is that several
of the LMC clusters with ages $\ge$100 Myr display [CO] indices 
much larger than those predicted by the models based on the Geneva's
tracks. 
In other words, a highly metallic
stellar population of 100--1000 Myr could have a [CO] similar to that of
a 10--100 Myr cluster of the same metallicity, as indeed predicted by
models including the whole AGB evolution (see Sect.~\ref{model_evol}).

Therefore, finding a galaxy with a very deep CO index 
does not necessarily imply
that its stellar population must be younger than 100 Myr, as sometimes assumed
in literature.

Fig.~\ref{figobs} also shows results from a wider set of data, 
including old
stellar systems (Galactic globular clusters and ellipticals)
and starburst galaxies.
In general, the only clear observational result is that objects with [CO]
significantly larger than 0.18 cannot be old stellar systems of (sub)solar
metallicities, but require younger stellar populations
or, alternatively, old stellar populations much more metallic 
than ellipticals.
Encouragingly, several starburst galaxies do indeed display CO
indices significantly larger than the above threshold 
but, in most cases, within the range covered by LMC clusters 
of ages $\la$10$^9$ yr
(see Fig.~\ref{figobs}).

On the other hand, however, other well studied starbursters have
values of [CO] lower than 0.18 and more similar to ellipticals and bulges.
This probably reflects metallicity variations, i.e.  weaker CO
features can be associated with young stellar systems of lower metallicities.

\section{Conclusions}
\label{conclusions}

Very large values of the CO index around 2.3 $\mu$m (i.e. 
[CO]$>$0.18, larger 
than those observed in the oldest, metal rich stellar systems) 
indicate that the near IR luminosity should be dominated by a relative 
young SP of RSG/AGB stars.
Any further attempt to better quantifying the age of this young SP 
in the range 10 Myr -- 1 Gyr is unreliable since the theoretical evolutionary
tracks are still too uncertain to properly describe 
the red supergiant and AGB evolution, especially at
sub--solar metallicities.

\begin{acknowledgements}
We would like to thank A. Bressan for helpful discussions and comments.
We are grateful to the referee, Y.D. Mayya, for comments and critics
which helped us to improve the quality of the paper.
This work was partly supported
by the Italian Ministry for University and Research (MURST) under grant
Cofin98-02-32.
\end{acknowledgements}

\end{document}